\documentclass[aps,showkeys,showpacs,superscriptaddress,a4paper,10pt]{revtex4-2}

\usepackage{amssymb}
\usepackage{amsmath}
\usepackage[dvips]{graphicx}
\usepackage[T2A]{fontenc}
\usepackage{graphicx}
\usepackage[dvips]{color}
\usepackage[breaklinks=true,colorlinks=true,linkcolor=blue,urlcolor=blue,citecolor=blue]{hyperref}

\begin{document}

\title[Higher order potential in the modified convective-viscous Cahn-Hilliard equation]{Higher order potential in the modified convective-viscous Cahn-Hilliard equation}

\author{P.~O.~Mchedlov-Petrosyan}
\author{L.~N.~Davydov}
\email[Corresponding author: ]{ldavydov@kipt.kharkov.ua}

\affiliation{National Science Center "Kharkiv Institute of Physics and Technology",   \\
1, Akademichna St., Kharkiv, 61108, Ukraine}

\author{O.~A.~Osmaev}
\affiliation{National Science Center "Kharkiv Institute of Physics and Technology",   \\
1, Akademichna St., Kharkiv, 61108, Ukraine}
\affiliation{Ukrainian State Railway Academy, 7, Feijerbakha Sq., Kharkiv, 61001, Ukraine}

\begin{abstract}
To describe highly heterogeneous systems using the Cahn-Hilliard equation, the standard form of the thermodynamic potential with a constant coefficient in the gradient term and a polynomial of the fourth degree may not be sufficient. The modification of the form of the thermodynamic potential with a polynomial of the sixth degree and the quadratic dependence of the coefficient at the gradient term is considered. Exact solutions in the form of a moving static wave and the conditions of their existence depending on the symmetry of the potential are obtained.
\end{abstract}


\maketitle

\renewcommand{\theequation}{\arabic{section}.\arabic{equation}}
\section{Introduction}\label{s1}

The present work is devoted to the study of the modified Cahn-Hilliard equation for the evolution of an order parameter $w$,
\begin{equation} \label{1.1} \frac{\partial w}{\partial t'} -2\bar{\alpha }w\frac{\partial w}{\partial x'} =\frac{\partial }{\partial x'} \left[M\frac{\partial }{\partial x'} \left(\bar{\mu }+\bar{\eta }\frac{\partial w}{\partial t'} \right)\right],   \end{equation}
\begin{equation} \label{1.2} \overline{\mu }=\overline{\varepsilon }\left[\frac{1}{2} \frac{dg}{dw} \left(\frac{\partial w}{\partial x'} \right)^{2} -\frac{\partial }{\partial x'} \left(g\left(w\right)\frac{\partial w}{\partial x'} \right)\right]+F_{5} (w).   \end{equation}
Here $\bar{\mu }$ is chemical potential, corresponding to double-well thermodynamic potential, containing $g\left(w\right)\left(\frac{\partial w}{\partial x} \right)^{2} $ term; $g\left(w\right)$ is a quadratic polynomial and $F_{5} \left(w\right)$ a quintic polynomial. To make clear the meaning of the above modification, we need to give some insight into the history and existing modifications of this equation. The Cahn-Hilliard equation \cite{1,2,3,4} is now a well-established model in the theory of phase transitions as well as in several other fields. The basic underlying idea of this model is that for inhomogeneous system, e.g. system undergoing a phase transition, the thermodynamic potential (e.g. free energy) should depend not only on the order parameter $w$, but on its gradient as well. The idea of such dependence was introduced already by Van der Waals \cite{5} in his theory of capillarity. For the inhomogeneous system the local chemical potential $\bar{\mu }$ is defined as variational derivative of the thermodynamic potential functional. If the thermodynamic potential is the simplest symmetric -- quadratic -- function of gradient this leads to the local chemical potential $\bar{\mu }$ which depends on Laplacian, or for the one-dimensional case -- on the second order derivative of the order parameter. The diffusional flux $J$ is proportional to the gradient of chemical potential $\nabla \overline{\mu }$; the coefficient of proportionality is called mobility $M$ \cite{6}. With such expression for the flux the diffusion equation, instead of usual second order equation, becomes a forth-order PDE for the order parameter $w$ (herein our notations differ from the notations in original papers):
\begin{equation} \label{1.3)} \frac{\partial w}{\partial t'} =\nabla \left[M\nabla \bar{\mu }\right], \end{equation}
\begin{equation} \label{1.4)} \bar{\mu }=-\bar{\varepsilon }^{2} \Delta w+F\left(w\right) .\end{equation}
Here $M$ is mobility, $\bar{\varepsilon }$ is usually presumed to be proportional to the capillarity length, and $F\left(w\right)=\frac{d\Phi \left(w\right)}{dw} $, where $\Phi \left(w\right)$ is homogeneous part of the thermodynamic potential. Originally $F\left(w\right)$ was taken in the form of the cubic polynomial (corresponding to the fourth-order polynomial for the homogeneous part of thermodynamic potential). The classic Cahn-Hilliard equation was introduced as early as 1958 \cite{1,2}; the stationary solutions were considered, the linearized version was treated and corresponding instability of homogeneous state identified. However, intensive study of the fully nonlinear form of this equation started essentially later \cite{7}. Now an impressive amount of work is done on nonlinear Cahn-Hilliard equation\textbf{\textit{,}} as well as on its numerous modifications, see \cite{3,4}. An important modification was done by Novick-Cohen \cite{8}. Taking into account the dissipation effects which are neglected in the derivation of the classic Cahn-Hilliard equation, she introduced the \textbf{\textit{viscous}} Cahn-Hilliard (VCH) equation
\begin{equation} \label{1.5} \frac{\partial w}{\partial t'} =\nabla \left[M\nabla \left(\bar{\mu }+\bar{\eta }\frac{\partial w}{\partial t'} \right)\right], \end{equation}
where the coefficient$\bar{\eta }$ is called viscosity. It was also noticed that VCH equation could be derived as a certain limit of the classic Phase-Field model \cite{9}. Later several authors considered the nonlinear \textit{convective} Cahn-Hilliard equation (CCH) in one space dimension \cite{10,11,12},
\begin{equation} \label{1.6} \frac{\partial w}{\partial t'} -\bar{\alpha }w\frac{\partial w}{\partial x'} =\frac{\partial }{\partial x'} \left(\frac{\partial \bar{\mu }}{\partial x'} \right) \end{equation}

Leung \cite{10} proposed this equation as a continual description of lattice gas phase separation under the influence of an external field. Similarly, Emmott and Bray \cite{12} proposed this equation as a model for the spinodal decomposition of a binary alloy in an external field \textit{E}. Witelski \cite{11} introduced the equation \eqref{1.6} as a generalization of the classic Cahn--Hilliard equation or as a generalization of the Kuramoto--Sivashinsky equation \cite{13,14} by including a nonlinear diffusion term. In \cite{10,11,12} and \cite{15,16} several approximate solutions and two exact static kink and anti-kink solutions were obtained. The `coarsening' of domains separated by kinks and anti-kinks was also discussed. Also the convective Cahn-Hilliard equation with cubic nonlinearity in the convective term was introduced in \cite{17,18}; however, in the present work we will consider only the Burgers-type nonlinearity, as in \eqref{1.6}. To study the joint effects of nonlinear convection and viscosity, Witelski \cite{19} introduced the convective-viscous-Cahn--Hilliard equation (CVCH) with a general symmetric double-well potential $\Phi \left(w\right)$:
\begin{equation} \label{1.7} \frac{\partial w}{\partial t'} -\bar{\alpha }w\frac{\partial w}{\partial x'} =\frac{\partial }{\partial x'} \left[M\frac{\partial }{\partial x'} \left(\bar{\mu }+\bar{\eta }\frac{\partial w}{\partial t'} \right)\right] ,  \end{equation}
\begin{equation} \label{1.8)} \overline{\mu }=-\overline{\varepsilon }^{2} \frac{\partial ^{2} w}{\partial x'^{2} } +\frac{d\Phi (w)}{dw} .   \end{equation}

It is worth noting that all his results, including the stability of solutions, were obtained without specifying a particular functional form of the potential. Thus, they are valid both for the polynomial and logarithmic potential. Also, with a constraint imposed on nonlinearity and viscosity, the approximate travelling-wave solutions were obtained. In \cite{20} for equation \eqref{1.7} with polynomial potential and the balance between the applied field and viscosity several exact single- and two-wave solutions were obtained.

Another line of modifications of the nonlinear Cahn-Hilliard equation which attracted much interest recently is the generalization of the form of the thermodynamic potential including higher-order polynomials $\Phi \left(w\right)$, non-constant coefficient $g\left(w\right)$ at $\left(\nabla w\right)^{2} $ and higher-order terms $\left(\Delta w\right)^{2}, \left(\nabla w\right)^{4} $, etc., resulting in the higher-order differential operator in equation. In \cite{21} Novick-Cohen noticed, that if the system as a whole is essentially inhomogeneous, the coefficient $g$ should be nonconstant. Such generalization of viscous CH equation was studied numerically in \cite{22} for the cubic polynomial in the potential. To the best of our knowledge the influence of non-constant $g$ on the movement of the phase transition fronts was not studied.

The sixth-order equations with nonconstant $g\left(w\right)$, and quintic polynomials where used systematically by Gomper and co-workers in modeling of Amphiphilic systems; this results in the sixth-order Cahn-Hilliard-type equation \cite{23,24,25,26}. This work was continued and elaborated by Pawlow and co-workers \cite{27,28,29}. They incorporated viscous effects associated with the rates of the order parameter and its spatial gradients. They considered initial/boundary value problem in 3D and obtained several rigorous mathematical results.

In the present work we will consider the fourth order convective-viscous CH equation with non-constant $g$, quintic polynomial, Burgers-type convective term, and constant mobility. The generalization for the sixth-order model, higher-order convective terms and non-constant mobility will be given elsewhere.

Our paper is organized as follows: in Section 2 we will obtain an exact travelling wave solution for the convective-viscous CH equation \eqref{1.1}, \eqref{1.2} with quntic polynomial $F_{5} \left(w\right)$. In Section 3 we study the dependence of solution on the parameters of the system. In Section 4 we will discuss our results.

\setcounter{equation}{0}
\section{Traveling wave solution}\label{s2}

We assume that the polynomial $F_{5} \left(w\right)$ has only three real roots, $\bar{a}_{1} <\bar{a}_{2} <\bar{a}_{3} $ corresponding to two minima and one maximum of the sixth-power two-wells polynomial in the thermodynamic potential. With such polynomial we rewrite \eqref{1.2} as
\begin{eqnarray} \label{2.1} & &{\overline{\mu }=-\overline{\varepsilon }\left[\frac{1}{2} \frac{\partial g}{\partial w} \left(\frac{\partial w}{\partial x'} \right)^{2} -\frac{\partial }{\partial x'} \left(g(w)\frac{\partial w}{\partial x'} \right)\right]} \nonumber \\ & &{+\overline{\rho }(w-\overline{a}_{1} )(w-\overline{a}_{2} )(w-\overline{a}_{3} )(w^{2} +b).} \end{eqnarray}
As it is usually presumed from symmetry considerations, we will use quadratic dependence of the coefficient $g$ on the order parameter, $g=\bar{\theta }w^{2} +\omega $; $b>0$.  Introducing non-dimensional order parameter $u=\frac{w}{\bar{a}_{3} } $, non-dimensional coordinate $x=\frac{x'}{X} $ and non-dimensional time $t=\frac{t'}{T} $, we rewrite equations \eqref{1.1} and \eqref{2.1} in non-dimensional form
\begin{equation} \label{2.2} \frac{\partial u}{\partial t} -2\alpha u\frac{\partial u}{\partial x} =\frac{\partial ^{2} }{\partial x^{2} } \left(\mu +\eta \frac{\partial u}{\partial t} \right) ,  \end{equation}
\begin{eqnarray} \label{2.3} & & {\mu =\left[-\theta u\left(\frac{\partial u}{\partial x} \right)^{2} -\left(\theta u^{2} +1\right)\frac{\partial ^{2} u}{\partial x^{2} } \right]} \nonumber \\ & &{+\rho \left(u-a_{1} \right)\left(u-a_{2} \right)\left(u-1\right)\left(u^{2} +\beta \right)}.  \end{eqnarray}

It is convenient to define $X=\sqrt{\bar{\varepsilon }\omega } ;\, \, \, T=\frac{\bar{\varepsilon }\omega }{M} $. We also introduced above the following notations: $\alpha =\overline{\alpha }\overline{a}_{3} \frac{\sqrt{\overline{\varepsilon }\omega } }{M} ;$ $\eta =\frac{\overline{\eta }M}{\overline{\varepsilon }\omega } ;$ $\theta =\frac{\overline{\theta }\overline{a}_{3}^{2} }{\omega } ;$ $\mu =\frac{\overline{\mu }}{\overline{a_{3} }} ;$ $\rho =\overline{\rho }\overline{a}_{3}^{4} $; $a_{1} =\frac{\bar{a}_{1} }{\bar{a}_{3} } ;\, \, \, a_{2} =\frac{\bar{a}_{2} }{\bar{a}_{3} } ;\, \, \, \beta =\frac{b}{\bar{a}_{3}^{2} } $.
Looking for travelling wave solution, we introduce $z=x-vt$. Equations \eqref{2.2}, \eqref{2.3} take form
\begin{equation} \label{2.4} -\frac{d}{dz} \left(vu+\alpha u^{2} \right)=\frac{d^{2} }{dz^{2} } \left(\mu -\eta v\frac{du}{dz} \right) ,  \end{equation}
\begin{eqnarray} \label{2.5} & &{\mu =\left[-\theta u\left(\frac{du}{dz} \right)^{2} -\left(\theta u^{2} +1\right)\frac{d^{2} u}{dz^{2} } \right]} \nonumber \\ & &{+\rho \left(u-a_{1} \right)\left(u-a_{2} \right)\left(u-1\right)\left(u^{2} +\beta \right)}.  \end{eqnarray}
Integrating \eqref{2.4} once we get
\begin{equation} \label{2.6} -\alpha \left(u^{2} +\frac{v}{\alpha } u+C_{1} \right)=\frac{d}{dz} \left(\mu -\eta v\frac{du}{dz} \right) .  \end{equation}

We are looking for solution which approaches values $u_{1} ,\, u_{2} $ at $\mp \infty $. For such solution the simplest proper Ansatz is
\begin{eqnarray} \label{2.7} & &{\frac{du}{dz} =\kappa \left[u-u_{1} \right]\left(u-u_{2} \right)=\kappa \left[u^{2} -pu+q\right]}, \nonumber \\ & &{p=u_{1} +u_{2} ;\, \, \, q=u_{1} u_{2} }. \end{eqnarray}
At $\pm \infty $ the right-hand side of \eqref{2.6} is zero; i.e. the left-hand side should be zero too. This means, that the polynomial in the left-hand side of \eqref{2.6} should coincide with the polynomial in the right-hand side of \eqref{2.7}. Setting $C_{1} =q$ and
\begin{equation} \label{2.8} v=-\alpha p=-\alpha \left(u_{1} +u_{2} \right) \end{equation}
we can rewrite the left-hand side of \eqref{2.6} as
\begin{equation} \label{2.9)} -\alpha \left(u^{2} +\frac{v}{\alpha } u+C_{1} \right)=-\alpha \left[u^{2} -pu+q\right]=-\frac{\alpha }{\kappa } \frac{du}{dz} .   \end{equation}
Integrating \eqref{2.6} once more, we get
\begin{equation} \label{2.10} \frac{\alpha }{\kappa } u+\mu -\eta v\frac{du}{dz} =C_{2} .   \end{equation}

It is convenient to introduce
\begin{equation} \label{2.11} \Phi \left(u\right)=-\left[\theta u\left(\frac{du}{dz} \right)^{2} +\left(\theta u^{2} +1\right)\frac{d^{2} u}{dz^{2} } \right]-\eta v\frac{du}{dz}  \end{equation}
and rewrite \eqref{2.10} as
\begin{eqnarray} \label{2.12} & &{\rho \left\{u^{5} -\left(\sigma +1\right)u^{4} +\left(\sigma +\zeta +\beta \right)u^{3} -\left[\beta \left(\sigma +1\right)+\zeta \right]u^{2} +\beta \left(\sigma +\zeta \right)u-\beta \zeta \right\}} \nonumber \\ & &{+\frac{\alpha }{\kappa } u+\Phi \left(u\right)=C_{2} \, ,} \end{eqnarray}
where we have denoted $a_{1} +a_{2} =\sigma ;\, \, a_{1} a_{2} =\zeta $. Using the Ansatz \eqref{2.7} all terms in \eqref{2.11} could be easily calculated:
\begin{eqnarray} \label{2.13}  & &{-\Phi \left(u\right)=3\theta \kappa ^{2} u^{5} -5\theta \kappa ^{2} pu^{4} +2\left[1+\theta p^{2} +2\theta q\right]\kappa ^{2} u^{3} }\nonumber \\ & &{+\left[\eta v-3p\kappa -3\theta \kappa pq\right]\kappa u^{2} +\left[-p\eta v+\kappa p^{2} +\left(\theta q+2\right)\kappa q\right]\kappa u+q\kappa \left(\eta v-\kappa p\right)} . \end{eqnarray}

Substitution of \eqref{2.13} into \eqref{2.12} yields
\begin{eqnarray} \label{2.14)}  & &{\left(\rho -3\kappa ^{2} \theta \right)u^{5} +\left[-\rho \left(\sigma +1\right)+5\theta \kappa ^{2} p\right]u^{4} }\nonumber \\ & &{+\left\{\rho \left(\sigma +\zeta +\beta \right)-2\kappa ^{2} \left[1+\theta p^{2} +2q\theta \right]\right\}u^{3} }\nonumber \\ & &{-\left\{\rho \left[\beta \left(\sigma +1\right)+\zeta \right]+\left[\eta v\kappa -3p\kappa ^{2} -3\theta \kappa ^{2} pq\right]\right\}u^{2} }\nonumber \\ & &{+\left[\rho \beta \left(\sigma +\zeta \right)+\frac{\alpha }{\kappa } +p\eta v\kappa -\kappa ^{2} p^{2} -\left(\theta q+2\right)\kappa ^{2} q\right]u=0}. \end{eqnarray}
Here $C_{2} $ was selected to eliminate the constant terms. Equating to zero coefficients at all powers of $u$ we obtain five constraints on the parameters
\begin{equation} \label{2.15} \kappa ^{2} =\frac{\rho }{3\theta },  \end{equation}
\begin{equation} \label{2.16} p=\frac{3}{5} \left(\sigma +1\right), \end{equation}
\begin{equation} \label{2.17} q=\frac{3}{4} \left(\zeta +\beta +\sigma \right)-\frac{1}{2} \left(\frac{1}{\theta } +p^{2} \right), \end{equation}
\begin{equation} \label{2.18} p\left(\alpha \eta \kappa +3\kappa ^{2} +3\theta \kappa ^{2} q\right)=\rho \left[\beta \left(\sigma +1\right)+\zeta \right] ,\end{equation}
\begin{equation} \label{2.19} \rho \beta \left(\sigma +\zeta \right)+\frac{\alpha }{\kappa } =p^{2} \left(\alpha \eta \kappa +\kappa ^{2} \right)+\left(\theta q+2\right)\kappa ^{2} q .\end{equation}

Using \eqref{2.8} we already eliminated $v$ from above equations. If constraints \eqref{2.15}-\eqref{2.19} and \eqref{2.8} are satisfied, the solutions of \eqref{2.7} are simultaneously solutions of \eqref{2.4}, \eqref{2.5}. Integrating \eqref{2.7} and taking the position of the maximal value of the derivative $\frac{du}{dz} $ as $z=0$, we get
\begin{equation} \label{2.20} u=\frac{u_{2} +u_{1} }{2} -\frac{u_{2} -u_{1} }{2} \tanh\left(\frac{1}{2} \kappa \left(u_{2} -u_{1} \right)\left(x-vt\right)\right) .\end{equation}
Despite the simple form of solution the dependence of its parameters on the parameters of the system is rather complicated; we analyze it in the next Section.

\setcounter{equation}{0}
\section{The parametric dependence of solution}\label{s3}

As it was shown in the previous Section, for \eqref{2.20} to be solution of the equation \eqref{2.4}-\eqref{2.5} six constraints should be satisfied. However, there are only four unknowns $\kappa ,\, q,\, p,$ and $v$. I.e., to have this form of solution two constraints should be imposed on the parameters of the system. For the general form of the polynomial the system of constraints is too complicated. To get some physical insight we will consider several special cases.
First, it's evident that for symmetric polynomial only static solution is possible: if $a_{1} =-1;\, \, a_{2} =0$ , i.e. $\zeta =0;\, \, \sigma +1=0$, from  \eqref{2.16} it follows $p=0,$ and from \eqref{2.8} $v=0$. The constraints \eqref{2.17} and \eqref{2.19} simplify drastically:
\begin{equation} \label{3.1} q=\frac{3}{4} \left(\beta -1\right)-\frac{1}{2\theta } , \end{equation}
\begin{equation} \label{3.2} \frac{\alpha }{\kappa ^{3} } =q\left(2+\theta q\right)+3\theta \beta . \end{equation}
The constraint \eqref{2.18} is satisfied automatically, both sides equal zero. Now, $p=0$ means $u_{1} =-u_{2} ;$ the solution \eqref{2.20} becomes
\begin{equation} \label{3.3)} u=-u_{2} \tanh\left(\kappa u_{2} x\right). \end{equation}
Thus the stationary solution is always symmetric, that is a kink for $\kappa <0$, and anti-kink for $\kappa >0$. While the constraint \eqref{2.15} allows both signs, \eqref{3.2} determines the sign of $\kappa $. It also shows that for such solution to exist there should be a definite relation between the applied field and parameters of the thermodynamic potential. It is interesting to consider the dependence of the right-hand side of \eqref{3.2} on $\theta $, i.e. on the ``measure of non-constancy of $g\left(w\right)$''. However, for $\beta >1$ the allowed values of $\theta $ are limited from above. Indeed, it follows from \eqref{3.1}
\begin{equation} \label{3.4} u_{2}^{2} =\frac{1}{2\theta } -\frac{3}{4} \left(\beta -1\right). \end{equation}
The right-hand side of \eqref{3.4} should be positive. It is fulfilled if $\beta \le 1$; if $\beta >1,$ there is an upper limit on $\theta $:
\begin{equation} \label{3.5} \theta <\frac{2}{3\left(\beta -1\right)} .\,  \end{equation}

Let us cosider the behavior of $\kappa ^{-3} $ as a function of $\theta $, see \eqref{3.2} and \eqref{3.1},
\begin{equation} \label{3.6} \frac{\alpha }{\kappa ^{3} } =\frac{3}{16\theta } \left\{\left[3\left(\beta -1\right)^{2} +16\beta \right]\theta ^{2} +4\theta \left(\beta -1\right)-4\right\}. \end{equation}
The right-hand side of \eqref{3.6} changes sign at $\theta =\theta _{1,2} $, where $\theta _{1,2} $ are the roots of quadratic equation
\begin{equation} \label{3.7} \theta ^{2} +\frac{4\left(\beta -1\right)}{\left[3\left(\beta -1\right)^{2} +16\beta \right]} \theta -\frac{4}{\left[3\left(\beta -1\right)^{2} +16\beta \right]} =0 ,\end{equation}
\begin{equation} \label{3.8} \theta _{1,2} =\frac{2}{\left(\beta +3\right)\left(3\beta +1\right)} \left\{-\left(\beta -1\right)\pm 2\left(\beta +1\right)\right\}. \end{equation}

From \eqref{3.8} the larger root is $\theta _{1} =\frac{2}{1+3\beta } $,  the lower root is negative; however $\theta >0$ per definition. I.e. the right-hand side of \eqref{3.6} is negative for $\theta <\theta _{1} $ and the sign of $\alpha $ should be opposite to the sign of $\kappa $. For $\theta >\theta _{1} $ the right-hand side of \eqref{3.6} is positive and the sign of $\alpha $ coincides with the sign of $\kappa $. For $\beta >1$ there is the upper limit on allowed values of $\theta $, see \eqref{3.5}; however, $\theta _{1} $ is less than this limit, i.e. the change of sign in \eqref{3.6} happens independent on the value of $\beta $. For $\theta =\theta _{1} $ it follows from \eqref{3.4} $u_{2} =1$; i.e. the boundary value of $u$ coincides with  the largest root of the polynomial in equation \eqref{2.3}, i.e. with the position of a minimum of two-well thermodynamic potential. On the other hand, if $\alpha =0$, $\theta =\theta _{1} $ is the only value of $\theta $ allowed by the constraint \eqref{3.6}. I.e. in the absence of the field the boundary values of $u$ coincide necessary with the positions of the minima of thermodynamic potential.
Also  \eqref{3.6} imposes the constraint on the ratio ${\alpha \mathord{\left/ {\vphantom {\alpha  \rho ^{\frac{3}{2} } }} \right. \kern-\nulldelimiterspace} \rho ^{\frac{3}{2} } } $
\begin{equation} \label{3.9} \frac{\alpha }{\rho ^{\frac{3}{2} } } =\frac{1}{16\sqrt{3} \left(\theta \right)^{\frac{5}{2} } } \left\{\left[3\left(\beta -1\right)^{2} +16\beta \right]\theta ^{2} +4\theta \left(\beta -1\right)-4\right\} .\,  \end{equation}
The ratio in the left-hand side of \eqref{3.9} reflects the balance between the applied field and polynomial part of the potential, necessary for the existence of exact static wave solution. While for $\theta =\theta _{1} $ the absolute value of the right-hand side crosses zero and then slightly increases, it evidently decreases as $\theta ^{-\frac{1}{2} } $ for large $\theta $. Naturally, with increase of $\theta $, i.e. increase of influence of derivative terms in the potential, the necessary influence of the field becomes weaker.

Now we consider the weakly asymmetric potentials. Introducing $\xi $ which could be positive or negative, $\left|\xi \right|\ll 1$, we will consider the small deviations of $a_{1} $ and $a_{2} $ from $-1$ and $0$, respectively, leading to different (up to the first order in $\xi $) expressions for the pairs $\sigma ,\zeta $.
\vspace{0.3cm}

I.   $a_{1} =-1+\xi ;\, \, a_{2} =0;\, \, \, \, \sigma =-1+\xi ;\, \, \zeta =0;$
\vspace{0.3cm}

II.  $a_{1} =-1;\, \, a_{2} =\xi ;\, \, \, \, \, \, \, \, \, \sigma =-1+\xi ;\, \, \zeta =-\xi ;$
\vspace{0.3cm}

III. $a_{1} =-1+\xi ;\, \, a_{2} =\xi ;\, \, \, \, \sigma =-1+2\xi ;\, \, \zeta =-\xi ;$
\vspace{0.3cm}

IV. $a_{1} =-1+\xi ;\, \, a_{2} =-\xi ;\, \, \, \, \sigma =-1;\, \, \zeta =\xi .$
\vspace{0.8cm}

I. For this case the constraints \eqref{2.16}-\eqref{2.18} take form
\vspace{0.3cm}

\begin{equation} \label{3.10} p=\frac{3}{5} \xi,  \end{equation}
\begin{equation} \label{3.11} q=\frac{3}{4} \left(\beta -1\right)-\frac{1}{2\theta } +\frac{3}{4} \xi , \end{equation}
\begin{equation} \label{3.12} \frac{\alpha \eta }{\kappa } =\frac{11}{4} \theta \beta +\frac{9}{4} \theta -\frac{3}{2} . \end{equation}
Constraint \eqref{3.12} is linking viscosity, applied field, and parameters of the potential, and naturally was not present for static solution. On the other hand, for the symmetric potential the general constraint \eqref{2.19} took the form \eqref{3.6}; the change of sign of the $\alpha \kappa $ took place at $\theta =\theta _{1} $, where $\theta _{1} $ is the larger root of the quadratic equation \eqref{3.7}. For the present case the form of the constraint \eqref{2.19} will be slightly different: correspondingly the coefficients of the quadratic equation may differ from the coefficients of \eqref{3.7} by the terms of order $\xi $. However in zero order the larger root $\theta _{1} $ still has to coincide with the root of
\begin{equation} \label{3.13)} 11\theta \beta +9\theta -6=0 ,\end{equation}
i.e. with the change of the sign of the right-hand side of \eqref{3.12}. This results in the compatibility condition
\begin{equation} \label{3.14)}  \beta =-3 . \end{equation}
However, $\beta $ is positive per definition, i.e. there is no solution for this type of asymmetry.
\vspace{0.8cm}

II.  $\sigma =-1+\xi ;\, \, \zeta =-\xi $.
\vspace{0.3cm}

For this case the constraints \eqref{2.16}-\eqref{2.18} take form
\begin{equation} \label{3.15} p=\frac{3}{5} \xi ,  \end{equation}
\begin{equation} \label{3.16} q=\frac{3}{4} \left(\beta -1\right)-\frac{1}{2\theta } +O\left(\xi ^{2} \right), \end{equation}
\begin{equation} \label{3.17)} \frac{\alpha \eta }{\kappa } =\frac{11}{4} \theta \left(\beta -1\right)-\frac{3}{2} .  \end{equation}
The right-hand side of the latter equation changes sign if $\theta $ equals 
\begin{equation} \label{3.18)} \theta =\frac{6}{11\left(\beta -1\right)} . \end{equation}

Again, in zero order in $\xi $ the parameter $\theta$ given by the latter equation should coincide with $\theta _{1} $, see \eqref{3.8}. For this case the compatibility condition is
\begin{equation} \label{3.19)} \beta =7 .\end{equation}
I.e. for the case II, we are considering now, the asymmetry results in non-zero velocity of the kink. The change of the sign of deviation from symmetry, $\xi $, changes the sign of the velocity
\begin{equation} \label{3.20)} v=-\alpha p=-\frac{3}{5} \alpha \xi . \end{equation}
Also the kink becomes asymmetric. To avoid confusion, we denote $u_{2} $ for the static kink \eqref{3.4} as $u_{s} $
\begin{equation} \label{3.21} u_{s}^{2} =\frac{1}{2\theta } -\frac{3}{4} \left(\beta -1\right) . \end{equation}
Looking for $u_{1} ,u_{2} $ in the form
\begin{equation} \label{3.22)} u_{1} =-u_{s} +\lambda _{1} \xi ;\, \, \, u_{2} =u_{s} +\lambda _{2} \xi , \end{equation}
we obtain from \eqref{3.15}-\eqref{3.16}
\begin{equation} \label{3.23)} \lambda _{1} +\lambda _{2} =\frac{3}{5} ;\, \, \, \lambda _{1} -\lambda _{2} =0 . \end{equation}
This results in
\begin{equation} \label{3.24)} u_{1} =-u_{s} +\frac{3}{10} \xi ;\, \, \, u_{2} =u_{s} +\frac{3}{10} \xi  \end{equation}

For the present case the constraint \eqref{2.19} takes the form
\begin{equation} \label{3.25} \frac{\alpha }{\kappa } =\frac{3\kappa ^{2} }{16\theta } \left\{\left[16\beta +3\left(\beta -1\right)^{2} \right]\theta ^{2} +4\left(\beta -1\right)\theta -4\right\}+O\left(\xi ^{2} \right). \end{equation}
As compared to \eqref{3.6} the latter equation shows only second order in $\xi $ corrections which cannot alter essentially the root of the polinomial in right-hand-side of \eqref{3.25}. Also this cannot change essentially the constraint imposed on the value of the ${\alpha \mathord{\left/ {\vphantom {\alpha  \rho }} \right. \kern-\nulldelimiterspace} \rho } ^{\frac{3}{2} } $ ratio. So we will not consider the analogues of \eqref{3.25}, i.e. the constraint \eqref{2.19} for other types of asymmetry. On the other hand, the constraints \eqref{2.16} -\eqref{2.18} are differently influenced by asymmetry and are considered for all cases.
\vspace{0.8cm}

III. $a_{1} =-1+\xi ;\, \, a_{2} =\xi ;\, \, \, \, \sigma =-1+2\xi ;\, \, \zeta =-\xi .$
\vspace{0.3cm}

For this case the constraints \eqref{2.16}-\eqref{2.18} yield
\begin{equation} \label{3.26)} p=\frac{6}{5} \xi , \end{equation}
\begin{equation} \label{3.27)} q=\frac{3}{4} \left(\beta -1\right)-\frac{1}{2\theta } +\frac{3}{4} \xi , \end{equation}
\begin{equation} \label{3.28)} \frac{\alpha \eta }{\kappa } =\frac{1}{4} \theta \left(11\beta -1\right)-\frac{3}{2} . \end{equation}
The right-hand-side of the latter equation changes sign for
\begin{equation} \label{3.29)} \, \, \theta =\frac{6}{11\beta -1} . \end{equation}
Equating again this $\theta $ to $\theta _{1} $, given by \eqref{3.8}, we obtain the compatibility condition for the present case
\begin{equation} \label{3.30} \beta =\frac{1}{2}. \end{equation}

The existence of positive $\beta $, given by \eqref{3.30}, ensures the existence of the traveling wave solution for this type of asymmetry. For this case the velocity is
\begin{equation} \label{3.31)} v=-\alpha p=-\frac{6}{5} \alpha \xi . \end{equation}
Looking again for $u_{1} ,u_{2} $ in the form
\begin{equation} \label{3.32} u_{1} =-u_{s} +\lambda _{1} \xi ,\, \, \, u_{2} =u_{s} +\lambda _{2} \xi , \end{equation}
where $u_{s} $ is given by\eqref{3.21}, we have
\begin{equation} \label{3.33} \lambda _{1} =\frac{1}{2} \left(\frac{6}{5} +\frac{3}{4u_{s} } \right) ,\, \lambda _{2} =\frac{1}{2} \left(\frac{6}{5} -\frac{3}{4u_{s} } \right) . \end{equation}
\vspace{0.8cm}

IV. $a_{1} =-1+\xi ;\, \, a_{2} =-\xi ;\, \, \, \, \sigma =-1;\, \, \zeta =\xi. $
\vspace{0.3cm}

The left-hand-side of \eqref{2.18} is proportional to \textit{p}. For the present case \eqref{2.16} yields $p\sim O\left(\xi ^{2} \right)$. However, the right-hand-side of \eqref{2.18} is $\rho \xi $, i.e. of the first order in $\xi $. So this constraint could not be satisfied. This means, that for this particular type of small asymmetry neither traveling, nor static solution is possible.

\setcounter{equation}{0}
\section{Discussion}\label{s4}

Summing up, we have obtained exact static and travelling wave solutions for the modified Convective-Viscous Cahn-Hilliard equation. The modification consists in the non-constant coefficient $g\left(w\right)$ at the $\left(\frac{\partial w}{\partial x} \right)^{2} $term and the sixth-power double-well polynomial in the thermodynamic potential. For such solutions to exist additional constraints should be imposed on the parameters of the system. While the form of solution is simple, the nonlinear algebraic system, linking parameters of solution to the parameters of the system, is quite complicated.

So we considered five cases: one, the completely symmetric quintic polynomial and, introducing a small parameter $\xi $, four different weakly asymmetric polynomials. For the symmetric potential only static solution is possible. Here is an interesting difference between the present equation and the standard form of the convective-viscous CH equation with the cubic polynomial: for the latter case the travelling-wave solution was obtained even for symmetric polynomial, if there is a special balance between he applied field and dissipation \cite{20}.

For the travelling wave the velocity of the front is always proportional to the deviation from the symmetry $\xi $ and the applied field $\alpha $. The non-dimensional parameter $\theta $  can be considered as a measure of ``non-constancy'' of the coefficient $g\left(w\right)$, see \eqref{1.2}, and according to the conjecture of \cite{21} as a measure of inhomogeniety. Returning to the dimensional parameters, we have $\theta =\frac{\bar{\theta }a_{3}^{2} }{\omega } $, where $a_{3} $ is the largest root of the quintic polynomial, and $\bar{\theta },\, \omega $ are coefficients of the quadratic polynomial $g\left(w\right)=\bar{\theta }w+\omega $. The dependence of the amplitude of the wave $u_{2} -u_{1} $ on $\theta $ is qualitatively different for two allowed cases of asymmetry. If $a_{1} =-1;\, \, a_{2} =\xi ;\, a_{3} =1$ (Case II), i.e. the positions of the stable states are fixed and only the unstable state deviates from zero, for $\beta =7$ the amplitude is

\begin{equation} \label{4.1)} u_{2} -u_{1} =2u_{s} =\sqrt{\frac{2}{\theta } -18}.  \end{equation}

So for small deviation $\xi $ the amplitude is independent on $\xi $ and the wave collapses to zero amplitude if $\theta $ approaches $\frac{1}{9} $from below; there are no solutions for larger $\theta $. If we consider $\theta $ as a measure of inhomogeneity, this means that for some level of disorder the constant velocity wave cannot exist. On the other hand, if $a_{1} =-1+\xi ;\, a_{2} =\xi ;\, a_{3} =1$ , the velocity is doubled, and for $\beta =\frac{1}{2} $ the amplitude is

\begin{equation} \label{4.2)} u_{2} -u_{1} =2u_{s} -\frac{3}{4u_{s} } \xi =\sqrt{\frac{2}{\theta } +\frac{3}{2} } +\frac{3}{4} \left[\frac{8\theta }{4+3\theta } \right]^{\frac{1}{2} } \xi . \end{equation}

There are no limitations on the value of $\theta $; for $\theta \gg 1$ the amplitude approaches the $\theta $-independent value; however it always depends on$\xi $:

\begin{equation} \label{4.3)} u_{2} -u_{1} \approx \sqrt{\frac{3}{2} } +\frac{3}{2} \sqrt{\frac{2}{3} } \xi \approx \sqrt{\frac{3}{2} } \left(1+\xi \right). \end{equation}

So in the latter case (when both the stable and unstable state deviate in the same direction from the symmetric position) the wave is less sensitive to the disorder, then in the former one. Interesting, when only stable state deviates (Case I), or when the stable and unstable states deviate from their equilibrium positions in opposite directions (Case IV), there is no solution for small deviations.

The general conclusion is that the velocity of the travelling waves, the amplitude of moving or static kinks, and even the very existence of solutions depend extremely strong on the asymmetry (or symmetry) of the sixth-power polynomial in the potential. Even small movement of the position of minima and maximum, i.e. the roots of the quintic polynomial in equation, results in changing the sign and absolute value of the velocity, and even the allowed intervals of the parameters.

{\small \topsep 0.6ex

}

\end{document}